\documentclass[aps, pra, twocolumn, superscriptaddress, nobalancelastpage,]{revtex4-2}

\pdfoutput=1

\usepackage[pdftex]{graphicx}
\usepackage{tikz}
\usepackage{hyperref}
\usepackage{mathtools}
\usepackage{amssymb}
\usepackage{wasysym}
\usepackage[utf8]{inputenc}
\usepackage[T1]{fontenc}
\usepackage{braket}
\usepackage[normalem]{ulem}
\usepackage{dsfont}
\usepackage{xcolor}
\usepackage{nicefrac}
\usepackage{enumerate}
\usepackage{float}
\makeatletter
\let\newfloat\newfloat@ltx
\makeatother
\usepackage{algorithm}
\usepackage{algpseudocode}
\usetikzlibrary{quantikz2}
\usepackage{rotating}
\usepackage{wrapfig}
\usepackage[caption=false]{subfig}
\usepackage{ragged2e}
\DeclareCaptionJustification{justified}{\justifying}
\usepackage{comment}
\makeatletter
\renewcommand*{\ALG@name}{Algorithm}
\makeatother

\newcommand{\St}{\mathrm{St}}
\newcommand{\Tet}{\mathrm{Tet}}
\newcommand{\Mor}{\mathrm{Mor}}

\newcommand{\gs}[1]{\mathrm{GS}_{#1}}
\newcommand{\MsqrtT}{\mathrm{H}_{\sqrt{\mathrm{T}}}}
\newcommand{\LMsqrtT}{\overline{\mathrm{H}}_{\sqrt{\mathrm{T}}}}
\newcommand{\MT}{\mathrm{H}_{\mathrm{T}}}

\DeclareMathOperator{\diag}{diag}

\newcommand{\I}{\mathrm{I}}
\newcommand{\X}{\mathrm{X}}
\newcommand{\Y}{\mathrm{Y}}
\newcommand{\Z}{\mathrm{Z}}
\newcommand{\T}{\mathrm{T}}
\newcommand{\Hg}{\mathrm{H}}
\newcommand{\Sg}{\mathrm{S}}
\newcommand{\sqT}{\sqrt{\mathrm{T}}}
\newcommand{\CX}{\mathrm{CX}}
\newcommand{\CZ}{\mathrm{CZ}}
\newcommand{\CCCZ}{\mathrm{CCCZ}}

\begin{document}

\title{Fault-tolerant $\ket{\sqT}$ state preparation and injection for more efficient fine-grained quantum circuit synthesis}
\author{Berat Yenilen}
\email{b.yenilen@fz-juelich.de}
\affiliation{\footnotesize Institute for Theoretical Nanoelectronics (PGI-2), Forschungszentrum Jülich, 52428 Jülich, Germany}
\affiliation{\footnotesize Institute for Quantum Information, RWTH Aachen University, 52074 Aachen, Germany}
\author{Markus Müller}
\affiliation{\footnotesize Institute for Theoretical Nanoelectronics (PGI-2), Forschungszentrum Jülich, 52428 Jülich, Germany}
\affiliation{\footnotesize Institute for Quantum Information, RWTH Aachen University, 52074 Aachen, Germany}
\email{-}
\author{Manuel Rispler}
\thanks{Current address: Alice \& Bob, 49 Bd du Général Martial Valin, 75015 Paris, France}
\affiliation{\footnotesize Institute for Theoretical Nanoelectronics (PGI-2), Forschungszentrum Jülich, 52428 Jülich, Germany}
\affiliation{\footnotesize Institute for Quantum Information, RWTH Aachen University, 52074 Aachen, Germany}

\begin{abstract}
Magic-state injection is a standard route 
to realize universal fault-tolerant quantum computation. 
Whereas the set of Clifford gates in combination with the non-Clifford T gate is a widely used universal gate set, extending the available set of non-Clifford primitives can reduce compilation overhead, provided that the additional primitives can be prepared fault-tolerantly with competitive resource costs and at sufficiently low logical noise rates. 
In this work, we introduce flag fault-tolerant protocols for preparing logical $\ket{\sqT}$ magic states on the 3D tetrahedral color code and its smaller morphed variant. 
Our simulations under circuit-level noise verify fault tolerance, quantify acceptance and logical error rates, and we reconstruct the effective logical channels of the corresponding circuits for gate injection via logical process tomography.
We find that access to $\sqT$ reduces the average space-time cost of synthesizing Haar-random single-qubit unitaries by approximately $20$--$30$\% relative to the Clifford$+\T$ gate set across practically relevant approximation regimes.
These results demonstrate how expanding the set of fault-tolerant non-Clifford primitives can improve computational efficiency and broaden the design space for universal quantum computation in the early fault-tolerant era.
\end{abstract}

\maketitle

\section{Introduction}
\label{section:introduction}

Quantum error correction (QEC) suppresses noise by encoding logical information into many physical qubits, thereby enabling quantum circuits that would otherwise be too deep to execute reliably~\cite{terhal_2015, campbell_review, gottesman_book}.
This noise protection, however, comes with substantial resource overhead~\cite{fowler_lattice_surgery_overhead_paper, litinski_2019} and, crucially, introduces new challenges for implementing logical gates in a fault-tolerant (FT) manner. Some logical operations are easier to implement fault-tolerantly, such as transversal gates that prevent error propagation within a code block; these often include some or all Clifford gates.
However, circuits composed solely of Clifford operations acting on stabilizer states can be efficiently simulated on a classical computer, a result known as the Gottesman--Knill theorem~\cite{gottesman_knill}.
Consequently, Clifford operations alone do not suffice for universal quantum computation: they must be supplemented by at least one non-Clifford operation~\cite{Kit97,nielsen_chuang}, such as the $\T$ gate in the Clifford$+ \T$ gate set or the Clifford$+$Toffoli gate set~\cite{eastin_2013_toffoli, jones_2013_toffoli}.
This requirement is difficult to reconcile with the most direct route to fault tolerance, as formalized by the Eastin--Knill theorem, which states that no QEC code can realize a universal logical gate set using only transversal gates~\cite{eastin_knill_theorem}.
As a result, at least one operation required for universality must be implemented using more elaborate FT techniques.

A standard way to circumvent this limitation is \emph{magic-state injection}~\cite{bk_2005, knill_2004}. 
In this approach, one prepares special ancilla states offline and consumes them via gate teleportation to enact logical operations that are not available transversally. 
The common framework is the Clifford$+ \T$ magic-state model~\cite{bk_2005, knill_2004, campbell_review, fowler_lattice_surgery_overhead_paper}: Clifford operations are implemented fault-tolerantly in a transversal manner, while the non-Clifford $\T$ gate (a $\pi/4$ $\Z$-rotation) is realized by injecting the magic state $\ket{\T}=\T\ket{+}$, where $\ket{+} =(\ket{0} + \ket{1})/\sqrt{2}$, often obtained through magic-state distillation~\cite{bk_2005, knill_2004, bravyi_haah} or by preparing it fault-tolerantly~\cite{chamberland_cross_2019, li_2015, gidney_cultivation, itogawaEvenMoreEfficient2024, vaknin_cultivation}.
Together with an entangling Clifford gate such as $\CX$, Clifford$+ \T$ is universal, and arbitrary unitaries can be approximated by sequences of Clifford and $\T$ gates.
While the Solovay--Kitaev theorem guarantees polylogarithmic scaling in the inverse synthesis precision~\cite{Kit97, nielsen_chuang}, practical and even asymptotically optimal synthesis methods for the Clifford$+ \T$ gate set have since been developed~\cite{selinger_ross, kliuchnikov_2023}, underscoring that compilation cost is a central driver of overhead in FT architectures.

Magic-state preparation and injection are central ingredients in fault-tolerant architectures, and they have therefore become key targets for near-term demonstrations of logical quantum computation. 
As the field moves from the Noisy Intermediate-Scale Quantum (NISQ) regime toward early FT quantum processors, alternative routes to universality—and low-overhead ways of realizing them—are likely to become increasingly important. 
Recent experiments have demonstrated several complementary ingredients: FT magic-state preparation and injection in trapped-ion systems~\cite{postlerDemonstrationFaulttolerantUniversal2022a,dasu_2025}; encoded magic-state preparation, cultivation, and injection on superconducting processors~\cite{ye2023logicalMagicStatePreparation, guptaEncodingMagicState2024,cultivation_experiment,kim2026magicStateInjectionIBM}; FT code switching ~\cite{Beverland2021Jun,butt_2024_fault-tolerant-code-switching,heussen_2025} as an alternative route to non-Clifford operations~\cite{experimental_code_switching,  daguerre2025codeSwitchingMagicStates}; and logical magic-state distillation in a reconfigurable neutral-atom and photonic architectures~\cite{souza2011experimentalMagicStateDistillation, rodriguez_2024, brown2023compilationMagicStateDistillation}. 
These developments make low-overhead protocols for preparing resource states particularly relevant, both as building blocks for large-scale architectures and as benchmarks for near-term logical non-Clifford operations.

Although one non-Clifford primitive is sufficient in principle, the cost of repeatedly producing and injecting $\ket{\T}$ states can dominate resource estimates~\cite{litinski_2019, campbell_review}.
This motivates enlarging the available logical instruction set with additional non-Clifford primitives~\cite{forest_2015, Mooney2021, amy_2024, dinh_2025} whenever they can be supplied fault-tolerantly.

In this work, we envision the logical data qubits as being encoded in the Steane code and introduce a flag-FT~\cite{Chao2018Aug, Chamberland2018Feb, gottesman_book} protocol for preparing the magic state $\ket{\sqT}=\sqT\ket{+}$.
Injecting this state into a Steane-encoded data qubit realizes a logical $\sqT$ gate and thereby extends the available fault-tolerant Clifford$+ \T$ gate set.
To prepare this state, we exploit the fault-tolerant logical $\T$ gate available in the $  [[15,1,3]] $ tetrahedral code and its smaller $[[10,1,2]]$ morphed variant~\cite{bombin_2006, morphing_quantum_codes}. 
Here, $[[n,k,d]]$ denotes a QEC code that encodes $k$ logical qubits into $n$ physical qubits and has distance $d$. 
The tetrahedral code corrects arbitrary single-qubit errors, while the morphed code uses fewer qubits but only detects such errors.

\begin{figure}[ht]
\captionsetup[subfigure]{}
\subfloat[$\T$-gate injection circuit.]{
\includegraphics[width=0.6\columnwidth]{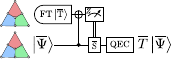}
}\newline
\subfloat[$\sqT$-gate injection circuit.]{
\includegraphics[width=0.9\columnwidth]{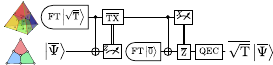}
}
\caption{Magic-state injection circuits for encoded qubits. The injected magic states are prepared fault-tolerantly: the $\ket{\sqT}$ state using the circuit in Fig.~\ref{fig:the_circuit}(a), and the $\ket{\T}$ state using the protocol of Ref.~\cite{chamberland_cross_2019}. The one-way transversal logical $\CX$ gate between the tetrahedral and Steane codes follows Ref.~\cite{heussen_2025}, while the noisy fault-tolerant QEC cycle is implemented using the method proposed in Ref.~\cite{poor_2025}.}
\label{fig:injection}
\end{figure}

The main contributions of this work are:
\begin{enumerate}
    \item We construct flag-FT preparation circuits for $\ket{\sqT}$ using the tetrahedral and the morphed code.
    \item We numerically characterize the preparation circuits under circuit-level depolarizing noise without idling, quantifying their acceptance probabilities, logical error rates (LERs), and physical resource requirements.
    \item We perform end-to-end circuit-level simulations to realize a fault-tolerant gate on a logical qubit using the circuit shown in Fig.~\ref{fig:injection}, including both fault-tolerant magic-state preparation and the subsequent injection circuit, and construct the resulting effective logical channel using quantum process tomography.
    \item We formulate a space-time cost model that incorporates the physical resources and logical noise associated with realizing a logical gate. 
    This model enables us to compare the physical space-time cost of realizing fault-tolerant gates on encoded qubits and to assess whether introducing the logical $\sqT$ primitive reduces the overall cost of approximating target single-qubit unitaries.
    \item Using this model, we perform noisy, cost-optimal single-qubit synthesis and compare the Clifford$+ \T$ and Clifford$+ \T + \sqT$ gate sets. 
    Accounting for both synthesis error and accumulated logical error, we find that access to $\sqT$ reduces the average space-time cost for Haar-random single-qubit unitaries by approximately $20\%$--$30\%$ across practically relevant approximation regimes.
\end{enumerate}

The remainder of the paper is organized as follows. Section~\ref{section:background} introduces the notation, noise model, and fault tolerance conventions used throughout, together with the necessary background on the relevant QEC codes and magic-state injection protocols. 
Section~\ref{section:fault-tolerant-sqrt-preparation} presents flag-FT protocols for preparing the logical $\ket{\sqT}$ magic state using the tetrahedral and the morphed code.
Section~\ref{section:preparation_numerics} numerically characterizes these preparation circuits in terms of their acceptance probabilities and LERs.
Section~\ref{subsection:qpt-ptm} describes the logical quantum process tomography of the circuits for state preparation and injection and presents the resulting logical noise characterization.
Section~\ref{section:noisy-database} introduces the expected space-time cost model and the method used for single-qubit gate synthesis. 
Finally, Section~\ref{section:comparison} compares the two gate sets and summarizes our main findings.

We note that, in parallel to the preparation of this work, Chen, da Silva Fonseca, and Sornborger independently proposed a protocol for cultivating magic states for the same $\ket{\sqT}$ resource state, based on phase-kickback checks in doubled color codes~\cite{Chen2026Jun}.

\section{Background and Notation}
\label{section:background}

In this section, we establish the notation employed throughout the manuscript and specify the fault tolerance model used in our analysis, followed by a concise review of the required background.

We use the standard definitions for the Pauli and Clifford operations, and define the following phase gates
\begin{align*}
\Sg &= \diag ( 1, e^{i \pi / 2 }) ,  \\  
\T  &= \diag ( 1, e^{i \pi/4} )  ,  \\
\sqT &= \diag ( 1, e^{i \pi/8 } ) .
\end{align*}
Throughout the main text, we use
\begin{equation*}
    \ket{\T}=\T\ket{+}, \qquad 
    \ket{\sqT}=\sqT \ket{+}, 
\end{equation*}
where $\ket{+} = (\ket{0} + \ket{1})/\sqrt{2}$. We also define
\begin{equation}
    \MT := \T \X \T^\dagger =  \Sg\X, \quad \MsqrtT := \sqT \X \sqT^\dagger = \T \X ,
\end{equation}
so that $\ket{\sqT}$ ($\ket{\T}$) is the $+1$ eigenstate of $\MsqrtT$ ($\MT$). 

We employ the same circuit-level depolarizing noise model throughout this work. 
The non-trivial Pauli components of each noise channel occur stochastically with total probability $p$, and each such occurrence is referred to as a \emph{fault}. 
The resulting deviation of the quantum state from its ideal evolution is referred to as an \emph{error}. 
Specifically,
\begin{enumerate}
    \item after each single-qubit gate and reset, and before each measurement, a single-qubit depolarizing channel with probability $p$ is applied;
    \item after each two-qubit gate, a two-qubit depolarizing channel with probability $p$ is applied; and
    \item idling errors are neglected.
\end{enumerate}
Concretely, the single-qubit depolarizing channel is
\begin{equation}
\label{eq:dep_channel_def}
\mathcal{D}_1(\rho) = (1-p)\rho + \frac{p}{3} \bigl( \X\rho\X + \Y\rho\Y+\Z\rho\Z \bigr),
\end{equation}
i.e. a fault is drawn from the (non-trivial) single qubit Paulis with probability $p/3$. Analogously, for the two-qubit channel a fault is drawn uniformly from the fifteen non-trivial two-qubit Pauli operators with probability $p/15$.

For a QEC code of distance $d=2t+1$, a circuit is \emph{fault tolerant} if any set of at most $t$ faults during its execution results in an output that remains correctable by the code and does not contain an undetected logical error~\cite{gottesman_book}. 
In particular, a single fault must not propagate into an uncorrectable multi-qubit error within a code block.

When a circuit uses post-selection, acceptance is determined by a prescribed set of measurement outcomes. 
Runs that satisfy these conditions are accepted, while all others are discarded. 
FT therefore requires that any accepted output produced in the presence of at most $t$ faults remain correctable, and that harmful fault patterns trigger rejection.
For the distance-three codes used below, $t=1$, this means that any accepted state must remain correctable after a single fault. 
For the distance-two morphed code, which is used as an error-detecting code, harmful single faults must instead be detectable and the corresponding logical states be rejectable by post-selection.
The following sections use these conventions to analyze magic-state preparation and injection circuits for the logical $\ket{\sqT}$ state.

\begin{figure*}[ht!]
    \centering
    \includegraphics[width=\linewidth]{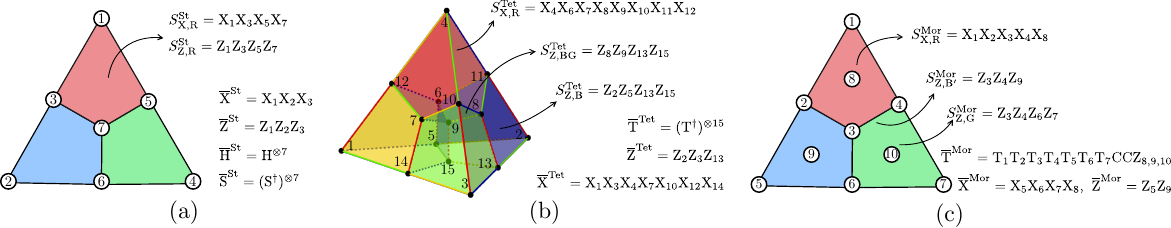}
\caption{
Representative quantum error-correcting codes used in this work.
(a) The Steane $[[7,1,3]]$ color code with face stabilizers $S^{\mathrm{St}}_{\X,c}$ and $S^{\mathrm{St}}_{\Z,c}$ for $c\in\{\mathrm{R,G,B}\}$ and transversal logical Clifford gates.
(b) The tetrahedral $[[15,1,3]]$ color code with cell $\X$-stabilizers $S^{\Tet}_{\X,c}$ ($c\in\{\mathrm{R,G,B,Y}\}$) and face $\Z$-stabilizers $S^{\Tet}_{\Z,f}$; it admits a transversal logical $\T$ gate.
(c) The morphed $[[10,1,2]]$ error-detecting code with pyramid-type 5-qubit $\X$-stabilizers and face-type $\Z$-stabilizers; it supports a fault-tolerant logical $\T$ implementation.
}
    \label{fig:color_codes}
\end{figure*}

\subsection{Steane code}

The Steane code is a CSS stabilizer code and the smallest non-trivial instance of the two-dimensional color-code family~\cite{Steane1996Jul,bombin_2006}. 
It encodes one logical qubit into seven physical qubits, can correct an arbitrary single-qubit error, and is therefore a $[[7,1,3]]$ code.

In Fig.~\ref{fig:color_codes}(a), we show the Steane code in its color-code representation, where physical qubits reside on vertices and $\X$- and $\Z$-type stabilizer generators are associated with the colored faces,
\begin{equation}
S^{\St}_{\X,c},\; S^{\St}_{\Z,c}, \quad c \in \{\mathrm{R,G,B}\}.
\end{equation}
Logical Pauli operators may be chosen transversally as
\begin{equation}
\overline{\X} = \X^{\otimes 7}, \quad 
\overline{\Z} = \Z^{\otimes 7},
\end{equation}
with equivalent minimum-weight representatives of weight three, establishing the code distance.

The main property we use is that the Steane code admits transversal Clifford gates. In particular,
\begin{equation}
\overline{\Hg} = \Hg^{\otimes 7}, \qquad 
\overline{\Sg} = (\Sg^\dagger)^{\otimes 7},
\end{equation}
and, due to its CSS structure, a transversal $\overline{\CX}$ can be implemented between two logical code blocks~\cite{terhal_2015}. 
Thus, Clifford operations are naturally FT in the Steane code. 
Consistent with the Eastin--Knill theorem, however, the Steane code does not provide a transversal universal gate set as it does not have a transversal $\T$ gate.

\subsection{Tetrahedral code}
\label{section:tetrahedral}

The three-dimensional tetrahedral color code~\cite{steane_qrm_codes,Bombin2007} is a CSS stabilizer code with parameters $[[15,1,3]]$. Geometrically, it can be understood as a color code defined on the cells and faces of a tetrahedron. 
A visual representation of the code is shown in Fig.~\ref{fig:color_codes}(b), where the vertices correspond to physical qubits.

The stabilizer generators of the tetrahedral code naturally decompose according to the geometry of the lattice. 
The $\X$-type stabilizer generators are supported on the four weight-8 cells of the tetrahedron.
The $\Z$-type stabilizers are associated with its weight-4 faces, from which we choose ten independent operators as generators.
We denote these stabilizers by
\begin{equation*}
    S^{\Tet}_{\X,c}, \quad S^{\Tet}_{\Z,f},
\end{equation*}
where $c \in \{ \mathrm{R, G, B , Y} \}$ labels the cells and $f \in \{ \mathrm{R, G, B, Y, RG, RB, RY, GB, GY, YB} \}$ labels the faces of the tetrahedron.
For example, $S_{\X, \mathrm{R}}^{\Tet}$ denotes the red-cell $\X$-stabilizer, which acts non-trivially on qubits $\{ 4,6,7,8,9,10,11,12 \}$. 
Similarly, $S_{\Z, \mathrm{BG}}^{\Tet}$ denotes the blue-green face $\Z$-stabilizer, acting on the face that is shared by the blue and the green cell, acting non-trivially on qubits $\{ 8, 9, 13, 15\}$.

The logical Pauli operators in the tetrahedral code admit transversal representatives,
\begin{equation*}
\overline{\X} = \X^{\otimes 15}, 
\qquad 
\overline{\Z} = \Z^{\otimes 15}.
\end{equation*}

Minimum-weight representatives of the logical operators can be obtained by multiplying the transversal operators with stabilizer generators. 
This yields face-like operators for the logical $X$,
\begin{equation*}
\overline{\X} = \X_1 \X_2 \X_3 \X_5 \X_{13} \X_{14} \X_{15},
\end{equation*}
and edge-like operators for the logical $Z$,
\begin{equation*}
\overline{\Z} = \Z_1 \Z_3 \Z_{14},
\end{equation*}
resulting in asymmetric distances $d_{\X}^{\Tet} = 7$ and $d_{\Z}^{\Tet} = 3$, for protection against phase- and bit-flip errors, respectively. 

A key feature of the tetrahedral code is that it is the smallest CSS code admitting a transversal implementation of the $\T$ gate~\cite{Koutsioumpas2022, bravyi_haah}:
\begin{equation*}
    \overline{\T} = (\T^\dagger)^{\otimes 15}.
\end{equation*}
However, the tetrahedral code does not support a transversal Hadamard gate and therefore does not realize a universal gate set transversally, as expected from the Eastin--Knill theorem~\cite{eastin_knill_theorem}.

\subsection{Morphed code}
\label{section:morphed}
The technique of code morphing is based on the idea that one can partially “unencode” a larger quantum code by applying the inverse encoding circuit of a smaller embedded subcode, thereby transforming the code into a different code with a similar structure.
By selectively unencoding one of the $\X$-type stabilizers of the tetrahedral code, one obtains a smaller $[[10,1,2]]$ stabilizer code, which for the purposes of this work we will refer to as \textit{the morphed code}~\cite{morphing_quantum_codes}. 
By morphing the tetrahedral code, we reduce the distance from three to two, such that the morphed code is an error-detecting rather than an error-correcting code, and its visual representation is shown in Fig.~\ref{fig:color_codes} (c).

The morphed code has three pyramid-type 5-qubit $\X$ stabilizers, denoted
\begin{equation*}
    S^{\Mor}_{\X,c}, \quad c \in \{\mathrm{R,G,B}\},
\end{equation*}
and six $\Z$ stabilizers. 
Three of the $\Z$ stabilizers are face-type weight-4 operators, denoted $S^{\Mor}_{\Z,c}$ for $c \in \{\mathrm{R,G,B}\}$, while the remaining three are weight-3 operators denoted $S^{\Mor}_{\mathrm{Z,R'}}$, $S^{\Mor}_{\mathrm{Z,B'}}$, and $S^{\Mor}_{\mathrm{Z,G'}}$ and are defined as 
\begin{align*}
    S^{\Mor}_{\mathrm{Z,R'}} &= \Z_3 \Z_6 \Z_8, \\
    S^{\Mor}_{\mathrm{Z,B'}} &= \Z_3 \Z_4 \Z_9, \\
    S^{\Mor}_{\mathrm{Z,G'}} &= \Z_2 \Z_3 \Z_{10}.
\end{align*}
One choice of minimum-weight logical Pauli representatives is
\begin{equation*}
    \overline{\X}^{\Mor} = \X_5 \X_6 \X_7 \X_8,
    \qquad
    \overline{\Z}^{\Mor} = \Z_5 \Z_9.
\end{equation*}
Similar to the tetrahedral code, the morphed code also has asymmetric distances $d_{\X}^{\Mor}=4$ and $d_{\Z}^{\Mor}=2$ and admits a FT implementation of a logical $\T$ gate,
\begin{equation}
    \overline{\T}^{\Mor}
    =
    \T_1 \T_2 \T_3 \T_4 \T_5 \T_6 \T_7 \, \mathrm{CCZ}_{8,9,10},
    \label{eq:morphed_T}
\end{equation}
where $\mathrm{CCZ}_{8,9,10}$ denotes a $\mathrm{CCZ}$ gate acting on qubits $8,9,10$. 
Despite not being transversal, this is FT in the sense that a failure of any individual constituent gate does not produce an undetected logical error. 
Although the morphed code has lower distance than the tetrahedral code, we show below that its smaller size and lower circuit overhead make it a useful alternative for FT $\ket{\sqT}$ preparation.

\subsection{Magic-state injection}
\label{section:magic_time}

\begin{figure}
    \centering
    \includegraphics[width=0.8\columnwidth]{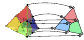}
    \caption{ The one-way transversal logical $\CX$ gate between the tetrahedral and Steane codes follows Ref.~\cite{heussen_2025}.}
    \label{fig:one_way_cx}
\end{figure}

We briefly recall the injection circuits used in our simulations.
A fault-tolerantly prepared Steane-encoded $\ket{\T}$ state can be consumed through the standard magic-state injection circuit shown in Fig.~\ref{fig:injection}(a), realizing a logical $\T$ gate.
Similarly, a fault-tolerantly prepared $\ket{\sqT}$ state in the tetrahedral code can be injected into a Steane-encoded data qubit using the circuit shown in Fig.~\ref{fig:injection}(b).
This realizes the logical $\sqT$ gate on the Steane code.
Here, we use the one-way logical transversal $\CX$ gate between the tetrahedral and Steane code, shown in Fig.~\ref{fig:one_way_cx}~\cite{heussen_2025}.
The gate is implemented transversally by applying physical $\CX$ gates between the Steane code and a suitable seven-qubit subset of the tetrahedral code.
It is one-way because this construction realizes a logical $\CX$ only for a fixed control-target qubit orientation.
In that orientation, the relevant stabilizers and logical Pauli operators propagate as required for a logical $\CX$, preserving fault tolerance through transversality~\cite{heussen_2025}.

\section{Fault-Tolerant $\ket{\sqT}$ Preparation}
\label{section:fault-tolerant-sqrt-preparation}

\begin{figure*}[t]
    \centering
    \captionsetup[subfigure]{font=normalsize}

    \subfloat[Fault-tolerant $\ket{\sqT}$ preparation circuit on the \textbf{tetrahedral code}]{
        \includegraphics[width=0.8\textwidth]{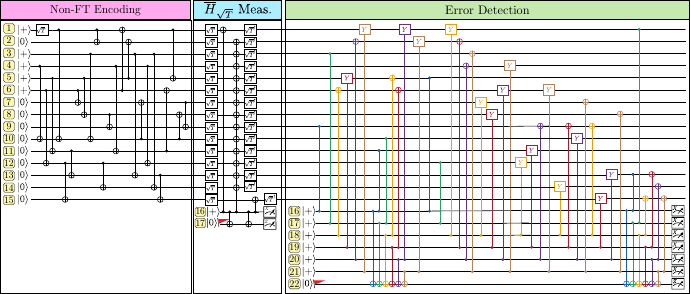}
    }
    \par\medskip

    \subfloat[Fault-tolerant $\ket{\sqT}$ preparation circuit on the \textbf{morphed code}]{
        \includegraphics[width=0.8\textwidth]{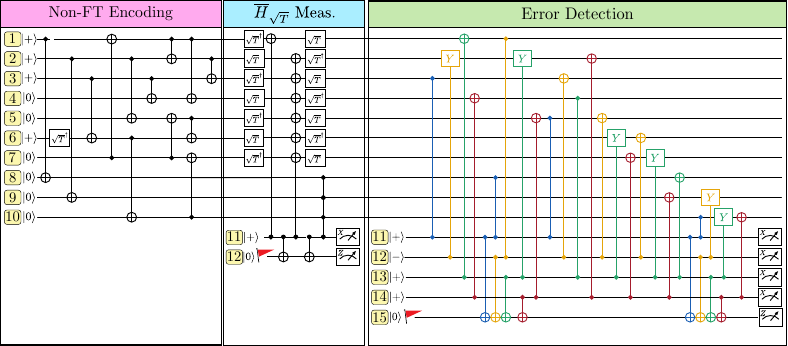}
    }

    \par\medskip

    \subfloat[CCCZ decomposition]{
        \includegraphics[width=0.8\textwidth]{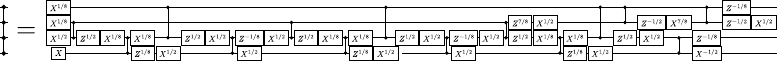}
    }

     \caption{ Quantum circuits for preparing the $\ket{\sqT}$ state on the (a) tetrahedral code and (b) morphed code, together with (c) the decomposition of the $\CCCZ$ gate required in the morphed-code circuit~\cite{cccz_decomposition}. Both preparation circuits consist conceptually of the same three components: (1) non-fault-tolerant unitary encoding, which encodes the physical $\ket{\sqT}$ state into the logical codespace of the code; (2) a flagged measurement of the logical operator $\LMsqrtT$, where the flag detects faults that could otherwise propagate into data errors containing a non-trivial logical component; and (3) fault-tolerant measurements of a reduced set of stabilizers chosen to detect the remaining harmful data errors arising from single faults that survive post-selection on the $\LMsqrtT$ measurement. Upon post-selecting on all the measurements being trivial, potential residual errors on the logical state are correctable (detectable) on the tetrahedral (morphed) code, as long as no more than a single fault occurs in the respective state preparation circuit.}
     \label{fig:the_circuit}
\end{figure*}

In this section, we present our FT protocols for preparing the $\ket{\sqT}$ state on the tetrahedral code and its morphed variant. 

\subsection{Preparation on the tetrahedral code}

We begin by observing that $\ket{\sqT}$ is the $+1$ eigenstate of the operator $\MsqrtT = \T \X$. 
On the tetrahedral code, this operator admits a transversal implementation:
\begin{equation*}
    \LMsqrtT = (\T^\dagger)^{\otimes 15} \X^{\otimes 15}, 
\end{equation*}
which can be rewritten as:
\begin{equation}
    \LMsqrtT = (\sqT^\dagger)^{\otimes 15} \X^{\otimes 15} (\sqT)^{\otimes 15}.
\end{equation}
This decomposition implies the following circuit identity:
\begin{center}
    \begin{quantikz}
        & \gate{\MsqrtT} & \\ 
        & \ctrl{-1}     & 
    \end{quantikz}
    =
    \begin{quantikz}
        & \gate{\sqT} & \targ{} & \gate{\sqT^\dagger} & \\ 
        &                  & \ctrl{-1} & &
    \end{quantikz}
\end{center}

Using this identity, we construct a FT $\ket{\sqT}$ preparation circuit inspired by Ref.~\cite{chamberland_cross_2019}. 
The full circuit is shown in Fig.~\ref{fig:the_circuit}(a) and consists of three components:

\begin{enumerate}
    \item \textbf{Non-FT encoding:} A unitary encoding circuit maps a physical $\ket{\sqT}$ state into the logical codespace of the tetrahedral code.

    \item \textbf{Flagged $\LMsqrtT$ measurement:} We fault-tolerantly measure the logical operator $\LMsqrtT$ using a flagged circuit.  Because the encoded state is ideally $\ket{\sqT}$, the $+1$ eigenstate of $\MsqrtT$, conditioning on the $+1$ outcome verifies the target logical state. Furthermore, the flag qubit detects dangerous errors that could otherwise turn into higher-weight errors on the first fifteen qubits.

    \item \textbf{Reduced FT stabilizer measurements:}
    Since the flagged $\LMsqrtT$ measurement already detects many of the dangerous single-fault events, it suffices to measure a subset of elements of the full stabilizer group that detects the remaining undetected single-fault events. We identify this subset by explicitly enumerating all single-fault events that survive the $\LMsqrtT$ measurement and selecting a subset of stabilizers that detects all remaining dangerous single-fault events while minimizing the total weight of the selected stabilizers. These stabilizers are then measured using shared flag qubits in a FT manner~\cite{liou_2025}.

\end{enumerate}

The protocol is a heralded, repeat-until-success preparation scheme. 
An attempt is accepted only when all measurement outcomes are trivial; if any nontrivial outcome is observed, the state is discarded and the preparation is restarted. 
We define the \emph{probability of success} as the acceptance probability of a single preparation attempt.

The scheme is FT in the sense that no single fault can produce an uncorrectable error on the data qubits. 
Upon acceptance, any error arising from a single physical fault is guaranteed to be correctable. 
Under the circuit-level depolarizing noise model introduced in Sec.~\ref{section:background}, in which faults occur independently with probability $p$ at each noisy circuit location, a logical error therefore requires at least two faults.
If one subsequently performs an ideal round of stabilizer measurements and decoding, the resulting state $\rho_{\mathrm{out}}$ has LER 
\begin{equation}
    \mathrm{LER_{out}}
    = 1 - \bra{\overline{\sqT}} \rho_{\mathrm{out}} \ket{\overline{\sqT}},
    \label{eq:ler_definition}
\end{equation}
and hence $\mathrm{LER_{out}}=O(p^2)$ in the low-noise regime. The expected quadratic behavior is confirmed numerically in Fig.~\ref{fig:numerics}.

\subsection{Preparation on the morphed code}

An analogous construction can be implemented on the morphed code, as shown in Fig.~\ref{fig:the_circuit}(b). 
The circuit follows the same three-stage structure: non-FT encoding, flagged $\MsqrtT$ measurement, and FT stabilizer measurements. 
As shown in Eq.~\eqref{eq:morphed_T}, the logical $\T$ implementation on the morphed code contains a $\mathrm{CCZ}_{8,9,10}$ gate. 
Since measuring $\LMsqrtT=\overline{\T}\,\overline{\X}$ requires its controlled implementation, controlling this $\mathrm{CCZ}$ factor introduces the measurement ancilla as an additional control, resulting in a $\CCCZ$ gate. Its decomposition into $\CZ$ gates and single-qubit rotations is shown in Fig.~\ref{fig:the_circuit}(c).

Since the morphed code is error-detecting, any single fault that results in an undetectable data error is filtered by post-selection. 
After acceptance, we perform an ideal round of stabilizer measurements and post-select on trivial outcomes. 
The LER is defined as in Eq.~\eqref{eq:ler_definition}, and the probability of success includes the probability of measuring the trivial syndrome in the ideal QEC round as well.

\section{Numerical Characterization of FT $\ket{\sqT}$ Preparation}
\label{section:preparation_numerics}

We now simulate the FT preparation circuits for both the tetrahedral and morphed codes. 
For the tetrahedral code, we employ Monte Carlo sampling, while for the morphed code we perform full density matrix simulations. 
For comparison, we also simulate FT $\ket{\T}$ preparation on the Steane code using full density matrix simulations, adopting the same definitions of acceptance probability and LER as in the preparation of the tetrahedral code.
We use the circuit from~\cite{chamberland_cross_2019} for the Steane $\ket{\T}$ preparation.

\begin{figure*}[ht!]
    \centering
    \includegraphics[width=0.9\linewidth]{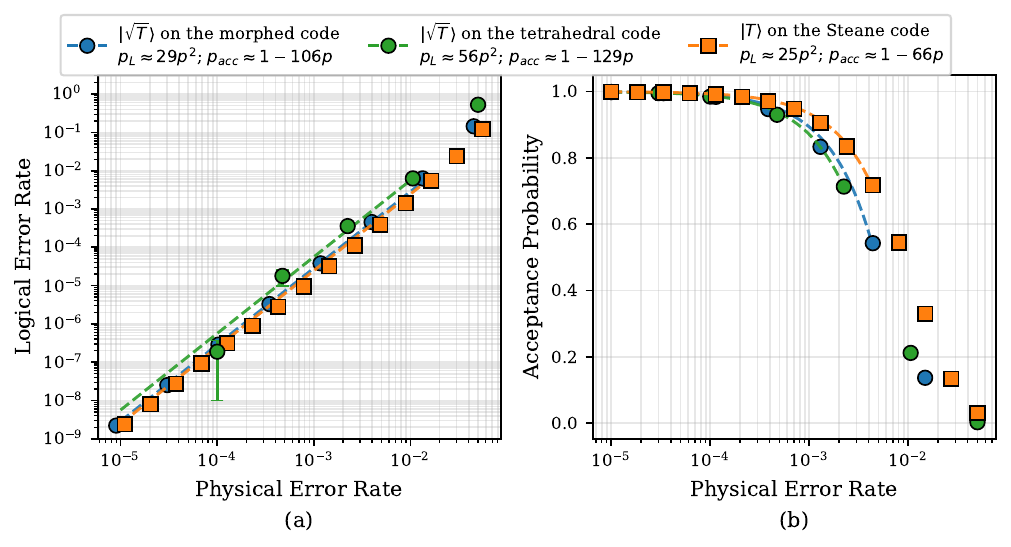}
    \caption{Numerical simulation of the expected performance of fault-tolerant $\ket{\sqT}$ preparation on the tetrahedral and morphed codes, compared with fault-tolerant $\ket{\T}$ state preparation on the Steane code. 
    (a) Logical error rate as a function of the physical error rate. 
    For the Steane and tetrahedral protocols, the logical error rate is computed after post-selection followed by an ideal round of stabilizer measurements and decoding. 
    For the error-detecting morphed code, we additionally post-select on trivial outcomes of the ideal stabilizer measurement outcomes; its acceptance probability therefore includes this additional post-selection step. 
    The curves are quadratic fits of the form $ A p^2$, with the estimated fit coefficients $A$ reported (rounded to integer values) in the legend.
    Error bars on the tetrahedral code results denote the standard error on the mean. 
    (b) Post-selection (acceptance) probability as a function of the physical error rate. We perform a linear fit to the acceptance probability and report it in the legend.}
    \label{fig:numerics}
\end{figure*}

The resulting LERs as a function of the physical error rate are shown in Fig.~\ref{fig:numerics}(a). 
All three protocols exhibit clear quadratic suppression in the low-$p$ regime and the quadratic fit results are reported on the legend.
The LER scales as approximately $29p^2$ and $56p^2$ when the $\ket{\sqT}$ state is prepared on the morphed and the tetrahedral code, respectively.
The LER scaling for preparing the $\ket{\T}$ state on the Steane code is found as $\approx 25p^2$.
This confirms that,  upon acceptance, logical errors arise from two or more physical faults, consistent with the expected behavior of the FT state preparation circuit.

Comparing the three schemes, the Steane $\ket{\T}$ preparation achieves the lowest LER across the considered range, while the tetrahedral $\ket{\sqT}$ preparation exhibits the highest LER, with the morphed protocol displaying a slightly worse error rate than the preparation protocol on the Steane code. 
Although the tetrahedral code has a larger distance than the morphed code, its preparation circuit involves a greater number of single- and two-qubit gates, leading to more fault locations. 
While single-fault events remain correctable by construction, the increased circuit volume raises the probability of higher-order fault combinations, resulting in a larger effective prefactor in the LER. 
This explains why the tetrahedral protocol displays a higher LER in our simulations despite its stronger underlying code properties.

Figure~\ref{fig:numerics}(b) shows the corresponding acceptance probabilities and the results from the respective linear fits are reported in the legend.
The Steane protocol has the highest acceptance rate, followed by the morphed protocol, while the tetrahedral protocol has the lowest acceptance probability over the entire range of physical error rates. 
Thus, the scheme on the tetrahedral code incurs both a stronger post-selection overhead and a larger residual LER under the considered noise model.
Importantly, however, this comparison concerns the preparation cost of an individual resource state. 
As we demonstrate in Sec.~\ref{section:comparison}, access to the additional $\sqT$ primitive can reduce the overall cost of synthesizing arbitrary single-qubit unitaries by enabling shorter and less resource-intensive decompositions, despite the higher cost of preparing the $\ket{\sqT}$ state.
Overall, the data illustrate the trade-off between circuit complexity, post-selection cost, and logical error suppression in FT state preparation.

\section{Logical Noise Characterization}

The simulations of the state preparation protocols above quantify the quality of the resource states prepared using the tetrahedral and the morphed codes.
In the following injection analysis, however, we consider only the tetrahedral-code preparation, for which the prepared state is injected
into a Steane-encoded data qubit.
We do not simulate injection from the morphed code, since the corresponding construction would place the logical data qubit in a distance-two error-detecting code.
To study the trade-off between preparation cost and computational power gained by augmenting the Clifford$+ \T$ gate set by a $\sqT$ gadget, we also require an effective logical noise model for the implemented gate itself.
This allows us to treat the complete procedure for preparing and injecting the state as an ideal logical gate followed by a noise channel and, for a given unitary decomposition, to estimate the logical noise accumulated across the full gate sequence. 
The LER of the resource state alone is insufficient for this purpose, since it does not specify the types of errors introduced by the injection gadget or how they propagate through subsequent gates. 
We therefore characterize the noisy logical operation by reconstructing its effective single-qubit channel. 
The same procedure is applied to the $\T$-gate injection and to the transversal Clifford gates used in the synthesis database; the corresponding additional data are reported in App.~\ref{app:ptm-pauli}.

\subsection{Simulations of $\sqT$ injection circuit}
\label{subsection:full-injection-simulations}

\begin{figure}
    \centering
    \includegraphics[width=\columnwidth]{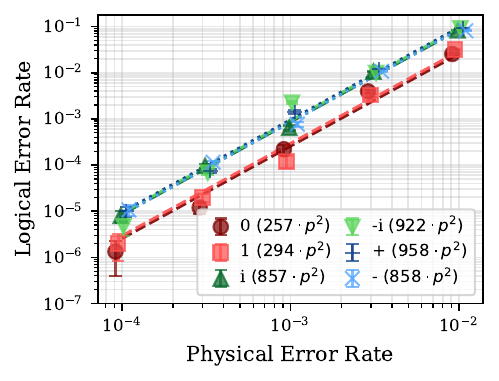}
    \caption{
    Six-state logical error rates for the noisy $\sqT$ injection circuit. 
    Each data set corresponds to one logical input state, and dashed lines show quadratic fits in the physical error rate $p$. 
    The approximately quadratic scaling confirms that the dominant accepted logical failures arise from second-order fault processes, while the variation among input states reveals anisotropic residual logical noise.
    } 
    \label{fig:qpt_ler_sqrtT}
\end{figure}

We simulate the full $\sqT$ injection circuit shown in Fig.~\ref{fig:injection} under the circuit-level depolarizing noise model introduced above. 
To probe the action of the noisy logical operation on the logical qubit Bloch sphere, we use the six logical cardinal input states
\begin{equation}
    \left\{
    \ket{0},\ket{1},\ket{+},\ket{-},\ket{\mathrm{i}},\ket{-\mathrm{i}}
    \right\},
\end{equation}
which correspond to the eigenstates of the three logical Pauli operators. 
After applying the noisy injection circuit, we perform an ideal round of QEC on the Steane code and decode the corrected output to the logical single-qubit statevector obtained in each Monte Carlo trajectory of a given noisy circuit realization. 
For an input state $\ket{\Psi}$ and target gate $G$, we compare the decoded output state $\ket{\Psi_{\mathrm{out}}}$ with the ideal target state
\begin{equation}
    \ket{\Psi_G}=G\ket{\Psi}.
\end{equation}
We define the LER for each input state as one minus the Monte Carlo average of the squared overlap with the target state,
\begin{equation}
    \operatorname{LER}_{G}(\Psi;p)
    =
1 - \mathbb{E}_{\mathrm{MC}}
    \left[
        \left|
        \braket{\Psi_G|\Psi_{\mathrm{out}} }
        \right|^2
    \right]
    \label{eq:qpt_LER}
\end{equation}
For the $\sqT$ injection circuit ($G=\sqT$),  Fig.~\ref{fig:qpt_ler_sqrtT} shows the resulting six-state LERs together with quadratic fits in the physical error rate $p$. 
All six inputs exhibit quadratic scaling in the low-noise regime, consistent with the FT suppression of single-fault logical failures. 
The fitted prefactors are state-dependent: the computational-basis inputs have smaller LERs than the equatorial inputs. 
This anisotropy indicates that a single averaged LER does not fully describe the implemented logical operation, motivating the analysis based on process tomography below.

We repeat the same characterization using six input states for the noisy $\T$ injection circuit and for the transversal $\Sg$ and $\Hg$ gates. 
For the $\T$ gate, the procedure is identical to the $\sqT$ case, with the expected state replaced by $\T\ket{\Psi}$. 
For $\Sg$ and $\Hg$, we use the corresponding noisy transversal Clifford implementation, followed by ideal error correction, decoding, and the LER evaluation as in Eq.~\eqref{eq:qpt_LER}. 
The extracted Pauli channel scalings for the $\T$ and $\Sg$ gates are reported in App.~\ref{app:ptm-pauli}, while the complete six-state LER data are provided in the accompanying Zenodo repository~\cite{zenodo_repository}.

\subsection{Pauli channel extraction via process tomography}
\label{subsection:qpt-ptm}
Throughout this subsection, we follow the Pauli transfer matrix (PTM) representation and conventions of Ref.~\cite{intro2gateset_tomography}.

The six-state data can be used to reconstruct the effective logical channel implemented by each noisy operation~\cite{nielsen_chuang}. 
For a target gate $G$, let $\mathcal{U}_{G}$ denote the ideal unitary channel and let $\widetilde{\mathcal{G}}(p)$ denote the simulated noisy logical implementation at physical error rate $p$. 
For each input state, we first decode the corrected Steane code output to a $2\times 2$ logical density matrix. 
The resulting pairs of logical input and output states are then used to reconstruct the single-qubit channel $\widetilde{\mathcal{G}}(p)$.

We represent the reconstructed channel using the Pauli-transfer-matrix representation (PTM).
With Pauli basis
\begin{equation}
    \{P_0,P_1,P_2,P_3\}=\{ \mathrm{ I,X,Y,Z} \},
\end{equation}
the Pauli transfer matrix of a single qubit channel $\mathcal{E}$ is defined as
\begin{equation}
    R_{ij}(\mathcal{E})
    =
    \frac{1}{2}
    \operatorname{Tr}
    \left[
        P_i\,\mathcal{E}(P_j)
    \right].
    \label{eq:ptm_def}
\end{equation}
To isolate the noise from the intended unitary action, we transform into the frame of the ideal gate and define the residual logical noise channel
\begin{equation}
    \mathcal{N}_{G}(p)
    =
    \widetilde{\mathcal{G}}(p)
    \circ
    \mathcal{U}_{G}^{-1}.
    \label{eq:residual_noise_channel}
\end{equation}
Equivalently, in the PTM representation and using the column-vector convention for Pauli coordinates,
\begin{equation}
    R_{\mathrm{err}}^{(G)}(p)
    =
    R(\widetilde{\mathcal{G}}(p))
    R(\mathcal{U}_{G})^{-1}.
    \label{eq:residual_ptm}
\end{equation}

We then test whether the residual logical noise is well approximated by a Pauli channel. 
Writing the residual PTM in block form as
\begin{equation}
    R_{\mathrm{err}}^{(G)}(p)
    =
    \begin{pmatrix}
        1 & (\vec{s}^{(G)}(p))^{T} \\
        \vec{t}^{(G)}(p) & A_{\mathrm{err}}^{(G)}(p)
    \end{pmatrix},
    \label{eq:ptm_block_form}
\end{equation}
trace preservation corresponds to $\vec{s}=\vec{0}$, while unitality corresponds to $\vec{t}=\vec{0}$. 
A Pauli channel is additionally diagonal in the lower-right $3\times 3$ block. 
We quantify the size of coherent or non-Pauli components using
\begin{equation}
    \Delta_{\mathrm{off}}^{(G)} (p)
    =
    \sum_{\substack{i,j\in\{X,Y,Z\}\\ i\neq j}}
    \left(A_{\mathrm{err},ij}^{(G)}(p)\right)^2.
    \label{eq:offdiag_metric}
\end{equation}
Small values of $\Delta_{\mathrm{off}}^{(G)} (p)$, together with small trace-preservation and unitality deviations, justify replacing the residual channel by its Pauli channel approximation.

\begin{figure}
    \centering
    \includegraphics[width=0.9\columnwidth]{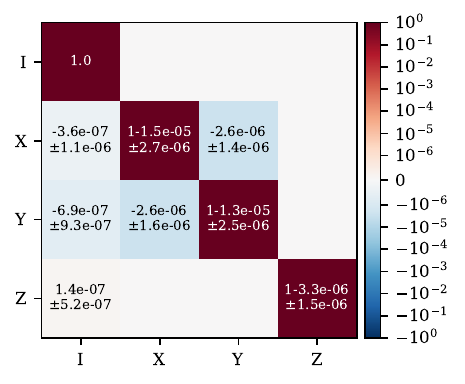}
    \caption{
    Pauli-transfer matrix of the residual error channel for the $\sqT$ injection circuit at physical error rate $p=10^{-4}$, after removing the ideal $\sqT$ action. 
    The displayed uncertainties are obtained from bootstrap resampling. 
    The first row and first column are consistent with trace preservation and unitality, respectively, while the off-diagonal entries of the $3\times 3$ Pauli block are small compared with the diagonal contractions.
    }
    \label{fig:QPT_PTM}
\end{figure}

Figure~\ref{fig:QPT_PTM} shows a representative residual PTM for the $\sqT$ injection circuit at $p=10^{-4}$. 
Within the bootstrap~\cite{efron_tibshirani_bootstrap} uncertainties, obtained by repeatedly resampling the Monte Carlo trajectories entering the reconstruction, the reconstructed channel is close to trace preserving and unital. 
Moreover, the off-diagonal entries in $A_{\mathrm{err}}^{(\sqT)}(10^{-4})$ are small compared with the diagonal entries. 
This supports the use of a Pauli channel approximation for the residual logical noise. 
The full set of reconstructed PTMs for the relevant operations, $\Sg$, $\Hg$, $\T$, and $\sqT$, at all simulated physical error rates is provided in the accompanying Zenodo repository~\cite{zenodo_repository}.

For a diagonal residual PTM, we denote its diagonal entries by
\begin{equation}
    \lambda_{P}^{(G)}(p)
    =
    A_{\mathrm{err},PP}^{(G)}(p),
    \qquad
    P\in\{\X,\Y,\Z\}.
\end{equation}
The corresponding Pauli channel approximation is parameterized by
probabilities $q_{P}^{(G)}(p)$ and acts as
\begin{equation}
    \rho
    \mapsto
    \sum_{P\in\{\I,\X,\Y,\Z\}}
    q_{P}^{(G)}(p)\,P\rho P.
\end{equation}
Using the PTM definition in Eq.~\eqref{eq:ptm_def}, its diagonal
entries satisfy (Walsh-Hadamard)
\begin{align}
    \lambda_{\X}^{(G)} (p)
    &= q_{\I}^{(G)} (p) + q_{\X}^{(G)} (p)
       -q_{\Y}^{(G)} (p) - q_{\Z}^{(G)} (p) ,\\
    \lambda_{\Y}^{(G)} (p)
    &= q_{\I}^{(G)} (p) - q_{\X}^{(G)} (p)
       + q_{\Y}^{(G)} (p) - q_{\Z}^{(G)} (p) ,\\
    \lambda_{\Z}^{(G)} (p)
    &= q_{\I}^{(G)}(p) - q_{\X}^{(G)}(p)
       - q_{\Y}^{(G)} (p) + q_{\Z}^{(G)}(p).
\end{align}
Together with
$\sum_{P}q_{P}^{(G)}=1$, inversion of these relations gives
\begin{align}
    q_{\I}^{(G)}(p) &=
\frac{
1+\lambda_{\X}^{(G)}(p)
+\lambda_{\Y}^{(G)}(p)
+\lambda_{\Z}^{(G)}(p)
}{4}, \\
q_{\X}^{(G)}(p) &=
\frac{
1+\lambda_{\X}^{(G)}(p)
-\lambda_{\Y}^{(G)}(p)
-\lambda_{\Z}^{(G)}(p)
}{4}, \\
q_{\Y}^{(G)}(p) &=
\frac{
1-\lambda_{\X}^{(G)}(p)
+\lambda_{\Y}^{(G)}(p)
-\lambda_{\Z}^{(G)}(p)
}{4}, \\
q_{\Z}^{(G)}(p) &=
\frac{
1-\lambda_{\X}^{(G)}(p)
-\lambda_{\Y}^{(G)}(p)
+\lambda_{\Z}^{(G)}(p)
}{4}.
    \label{eq:pauli_probs_from_ptm}
\end{align}
We apply this extraction to each bootstrap sample and use the resulting distribution to estimate error bars on the Pauli probabilities.

\begin{figure}
    \centering
    \includegraphics[width=0.95\columnwidth]{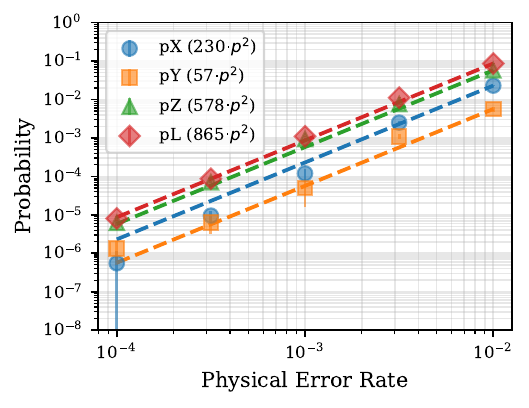}
    \caption{
    Pauli channel probabilities for the residual noise of the $\sqT$ injection circuit, obtained after removing the ideal $\sqT$ action. 
    Error bars are obtained by bootstrap resampling of the reconstructed PTMs. 
    The residual channel is dominated by a $\Z$-type component, followed by smaller $\X$ and $\Y$ components.
    }
    \label{fig:QPT_pauli_probs}
\end{figure}

Fig.~\ref{fig:QPT_pauli_probs} shows the extracted Pauli probabilities for the $\sqT$ injection circuit. 
The fitted quadratic scalings are 
\begin{equation}
    q_{\X}^{(\sqT)}(p) = 230p^2,\ 
    q_{\Y}^{(\sqT)}(p) = 57p^2,\ 
    q_{\Z}^{(\sqT)}(p) = 578p^2.
    \label{eq:sqrtT_pauli_fits}
\end{equation}

The extracted coefficients show that the residual logical noise is substantially biased toward $\Z$-type errors. 
This is consistent with the mechanism identified in ~\cite{Fazio2025Jun_injection_bias}, where magic-state injection can transform noise on the injected resource into phase noise on the computational logical qubit, producing a $\Z$-biased logical error model even when the underlying physical noise is not $\Z$-biased. 
The particularly small $\Y$-component can be understood from the subsequent ideal syndrome extraction: since $\X$- and $\Z$-type syndromes are corrected sequentially, a logical $\Y$-type residual, which contains both $\X$ and $\Z$ components, can be partially resolved into the corresponding single-Pauli correction sectors. 
This shifts part of the apparent $\Y$-weight into the extracted $\X$- and $\Z$-components, leaving $q_{\Y}^{(\sqT)} (p)$ comparatively suppressed.

To obtain a simpler noise model that can be used in the later compilation analysis, we replace the anisotropic residual Pauli channel by the single-qubit depolarizing channel defined in Eq.~\eqref{eq:dep_channel_def}, with its error probability chosen conservatively as
\begin{equation}
    q^{(G)}_{\mathrm{dep}}(p)
    =
    3\max\left\{
        q_{\X}^{(G)}(p),
        q_{\Y}^{(G)}(p),
        q_{\Z}^{(G)}(p)
    \right\}.
    \label{eq:depol_upper_bound_rule}
\end{equation}
This choice upper-bounds each non-identity Pauli probability by the corresponding depolarizing probability $q^{(G)}_{\mathrm{dep}}/3$. 
For the $\sqT$ injection circuit in Fig.~\ref{fig:injection}(b), Eq.~\eqref{eq:sqrtT_pauli_fits} gives
\begin{equation}
    q^{(\sqT)}_{\mathrm{dep}}(p)
    \simeq
    1734p^2.
    \label{eq:sqrtT_depol_fit}
\end{equation}

We perform the same reconstruction, Pauli channel extraction, and procedure for obtaining a depolarizing upper bound for the $\T$ injection circuit and for the transversal $\Sg$ and $\Hg$ gates. 
The resulting coefficients are summarized in Tab.~\ref{tab:primitive-noise-costs}.

\section{Noisy Cost-Aware Synthesis of Arbitrary Single-Qubit Operations}
\label{section:noisy-database}

We now describe how the logical primitives characterized above are used in the single-qubit synthesis benchmarks. 
The goal is to compare two gate sets,
\begin{align}
    \gs{1} &= \{ \mathrm{ H,S,S^\dagger,T,T^\dagger} \}, \\
    \gs{2} &= \gs{1} \cup\{\sqT,\sqT^{\dagger}\},
    \label{eq:gs_def}
\end{align}
while accounting for both the physical space-time cost of each logical primitive and the logical noise accumulated by a synthesized sequence.
The daggered gates are implemented using the complex-conjugate
preparation and injection circuits. 
Under the depolarizing noise model, these circuits have the same cost and effective depolarizing noise as their un-daggered counterparts; hence, we assume $C_{G^\dagger}=C_G$ and $q^{(G^\dagger)}_{\mathrm{dep}} (p) =q^{(G)}_{\mathrm{dep}} (p) $.

\subsection{Space-time cost model}

We quantify the implementation cost of a circuit by its physical space-time volume~\cite{liou_2023}. 
This quantity measures how many physical qubits are occupied, and for how long, during the execution of a circuit block. 
Concretely, for a circuit block using $n_q$ physical qubits over $n_t$ time steps, we assign the volume
\begin{equation}
    V = n_q n_t.
\end{equation}
For circuits composed of several blocks, we compute the volume of each block and sum the contributions. 
Throughout this work, we assume that qubits can be initialized and measured in either the $\X$ or $\Z$ basis, that qubit connectivity is all-to-all, and that $\CX$ gates acting on disjoint pairs of qubits can
be performed in parallel.
For example, in Fig.~\ref{fig:the_circuit}(a), the encoding block uses $15$ qubits and its $\CX$ gates can be scheduled in five time steps, giving a space-time volume of $75$.

For a post-selected logical primitive $G$, the relevant cost is the expected space-time volume required to obtain one accepted implementation,
\begin{equation}
    C_G(p)
    =
    \frac{V_G}{P_{\mathrm{acc}}^{(G)}(p)}.
    \label{eq:expected_cost}
\end{equation}
Here $V_G$ is the space-time volume of the corresponding implementation circuit and $P_{\mathrm{acc}}^{(G)}(p)$ acceptance probability obtained from the fit to the simulated acceptance rates at physical error rate $p$. 
The factor $1/P_{\mathrm{acc}}^{(G)}(p)$ is the expected number of attempts
required to obtain one accepted implementation.
For transversal Clifford gates, which do not involve post-selection in our model, we assign a fixed space-time cost equal to one transversal single-qubit layer on the Steane code.

The logical-noise coefficients and expected costs used in the compilation analysis are summarized in Tab.~\ref{tab:primitive-noise-costs}. 
The expected costs include the overhead from repeating post-selected FT circuits for state preparation until acceptance. 
As a result, they acquire a weak dependence on the physical error rate $p$, through the corresponding acceptance probabilities. 
In the physical error regime considered below, $p\in[10^{-5},10^{-3}]$, this dependence is negligible compared with the differences between primitive costs. 
We therefore neglect this dependence in the synthesis database and use the normalized cost in Tab.~\ref{tab:primitive-noise-costs}.

\begin{table}[ht!]
\centering
\begin{tabular}{c|c|c|c}
\hline
Primitive $G$ 
& $q_G^{\mathrm{dep}}(p)$ 
& $C_G(p)$ 
& Normalized cost \\
\hline
Transversal Clifford
& $21p^2$ 
& $7$ 
& $\approx 0.02$ \\
$\T$ injection 
& $951p^2$ 
& $400 / (1-66p)$ 
& $1$ \\
$\sqT$ injection 
& $1734p^2$ 
& $ 980 / (1-129p)$ 
& $ \approx 2.45$ \\
\hline
\end{tabular}
\caption{
Effective logical noise and expected space-time costs for the primitive operations used in the compilation analysis. 
We assume that the $\T$ and $\sqT$ gates are implemented using the injection circuits shown in Fig.~\ref{fig:injection}.
The required magic states are prepared fault-tolerantly: the $\ket{\T}$ state is prepared in the Steane code using the protocol of Ref.~\cite{chamberland_cross_2019}, and the $\ket{\sqT}$ state is prepared in the tetrahedral code using the circuit in Fig.~\ref{fig:the_circuit}(a).
The logical noise of each primitive is modeled as an ideal gate followed by a depolarizing channel $\mathcal{D}_1 (q_G^{\mathrm{dep}}(p))$. 
The expected costs include post-selection overheads through the fitted acceptance probabilities, while the normalized costs report the leading-order ratios relative to one accepted $\T$ injection.
}
\label{tab:primitive-noise-costs}
\end{table}

\subsection{Database construction}

We construct a database of cost-optimal decompositions of the single-qubit unitaries generated by each gate set up to a maximum sequence cost $C_{\max}$, using a Dijkstra-style tree search, following the strategy of Algorithm~1 in Ref.~\cite{Mooney2021}. 
The identity forms the root of the tree, and each edge appends one allowed elementary operation.
A node therefore corresponds to the gate sequence along the path from the root to that node, and to the unitary implemented by that sequence.
The search expands nodes in increasing order of total sequence cost, so the first sequence found for a distinct unitary is cost-optimal with respect to the assigned primitive costs. 
The search terminates when the lowest-cost unprocessed node has cost greater than $C_{\max}$.
The resulting finite database is later queried to approximate arbitrary target unitaries.

To represent single-qubit unitaries geometrically, we ignore global phase and work in $\mathrm{PSU(2)}$. 
For each unitary $U$, we choose the representative
\begin{equation}
    U \equiv \exp\!\left[i(\alpha \X +\beta \Y +\gamma \Z)\right],
\end{equation}
and associate to it the Pauli-vector coordinate
\begin{equation}
    \vec{v}_U = (\alpha,\beta,\gamma).
\end{equation}
Two unitaries are treated as identical when the Euclidean distance
between their Pauli-vector coordinates is smaller than $10^{-8}$.

This vector lies in a ball of radius $\pi/2$. 
Its direction specifies the rotation axis, while its norm specifies the rotation angle. 
For example, the identity is represented by $\vec{v}_I=(0,0,0)$, while a $\Z$ rotation is represented by $\vec{v}_{\Z} = (0,0,\pi/2)$ (up to an irrelevant global phase).

For each gate set, we generate a database as follows. 
For $\gs{1}$, the branching operations are the $24$ single-qubit Clifford operations together with $\T$ and $\T^\dagger$. 
For $\gs{2}$, we use the same operations and additionally include $\sqT$ and $\sqT^{\dagger}$. 
Each operation is assigned the normalized cost listed in Tab.~\ref{tab:primitive-noise-costs}. 
The database construction then proceeds as follows:
\begin{enumerate}
    \item Initialize the sequence tree with the identity as the root node.
    \item Store all current leaf nodes in a minimum heap ordered by total sequence cost.
    \item Pop the lowest-cost leaf node from the heap and compute the Pauli-vector coordinate of the unitary represented by that node.
    \item If this coordinate has not yet been encountered, store the corresponding sequence in the database and mark the coordinate as visited.
    \item Generate child nodes by appending each allowed operation in the gate set.
    \item Add the new child nodes to the heap unless their accumulated unitaries have already been reached at lower cost.
    \item Repeat until a chosen maximum sequence cost is reached.
\end{enumerate}
The databases used in this work are generated with maximum sequence costs $17$ for $\gs{1}$ and $13$ for $\gs{2}$.
The resulting database stores, for each retained Pauli-vector coordinate, the lowest-cost sequence that realizes the corresponding unitary, up to the numerical tolerance defined above, under the specified cost model.
As in Ref.~\cite{Mooney2021}, we store the database in a nearest-neighbor data structure so that target unitaries can be queried geometrically in the Pauli-vector representation.

\subsection{Query procedure under a total error budget}
\label{subsection:query-total-error}

Approximating a target unitary more accurately generally requires a
longer gate sequence, which reduces the synthesis error but increases
both the space-time cost and the accumulated logical noise.
The database query accounts for this trade-off by searching for the
lowest-cost sequence that satisfies a prescribed total error budget.
The input to the query is a target unitary $U$, a physical error rate $p$, and a total error budget $\epsilon_{\mathrm{tot}}$. 
We first compute the Pauli-vector coordinate $\vec{v}_U$ of the target unitary and retrieve the $5000$ nearest stored database entries using Euclidean distance in the Pauli-vector representation. 
Each retrieved entry corresponds to a candidate compiled unitary $\widetilde{U}_j$, together with an assigned space-time cost and gate counts
\begin{equation}
    n_{\mathrm{C},j},\qquad n_{\T,j},\qquad n_{\sqT,j}.
\end{equation}
Here, $n_{\mathrm{C},j}$ is the number of Clifford layers, while $n_{\T,j}$ and $n_{\sqT,j}$ count the number of $\T$ or $\T^\dagger$ injections and $\sqT$ or $\sqT^{\dagger}$ injections, respectively.
Each injection count refers to one complete injection circuit shown in Fig.~\ref{fig:injection}, including all measurement-conditioned corrections. 
In particular, any $\T$ ($\Sg$) correction required to realize $\sqT$ ($\T$) is already included in the cost and logical noise assigned to the corresponding $\sqT$ ($\T$) injection and is not counted separately.

For each candidate, we first compute the synthesis error between the target unitary $U$ and the noiseless compiled unitary $\widetilde{U}_j$. 
Throughout this work, we use the diamond-norm convention~\cite{nielsen_chuang} and for single-qubit unitaries, this can be calculated using
\begin{equation}
    \epsilon_{\mathrm{synth},j}
    =
    \left\|
        \mathcal{U}
        -
        \widetilde{\mathcal{U}}_j
    \right\|_{\diamond}
    =
    2
    \sqrt{
        1-
        \frac{
            \left|
            \operatorname{Tr}(U^\dagger \widetilde{U}_j)
            \right|^2
        }{4}
    }.
    \label{eq:unitary_diamond_distance}
\end{equation}

We then estimate the logical error accumulated by the noisy implementation of the candidate sequence. 
Each primitive gate $G$ is modeled as the ideal operation followed by a depolarizing channel,
\begin{equation}
    \widetilde{\mathcal{G}}
    =
    \mathcal{D}_1 ( q^{(G)}_{\mathrm{dep}}(p) ) 
    \circ
    \mathcal{U}_G,
    \qquad
    q^{(G)}_{\mathrm{dep}}(p)=c_Gp^2,
    \label{eq:noisy_primitive_model}
\end{equation}
where the coefficients $c_G$ are listed in Tab.~\ref{tab:primitive-noise-costs}. 
Here, $\mathcal{D}_1 (q^{(G)}_{\mathrm{dep}}(p))$ denotes the single-qubit
depolarizing channel defined in Eq.~\eqref{eq:dep_channel_def}, which
models the residual logical noise following the ideal action of the
primitive gate $G$.

Since depolarizing noise is invariant under unitary conjugation, the depolarizing channels associated with the gates in a compiled sequence can be commuted to the end of the ideal circuit. 
In the low-noise regime considered here, the accumulated depolarizing parameter is therefore additive to leading order in $p^2$. 
For a candidate sequence with gate counts $n_{\mathrm{C},j}$, $n_{\T,j}$, and $n_{\sqT,j}$, we use
\begin{equation}
    q_{\mathrm{acc},j}(p)
    =
    \left(
        n_{\mathrm{C},j}c_{\mathrm{C}}
        +
        n_{\T,j}c_\T
        +
        n_{\sqT,j}c_{\sqT}
    \right)p^2.
    \label{eq:pacc_second_order}
\end{equation}
The omitted corrections are $O(p^4)$ and are negligible for the physical error rates considered in this work, $p\in[10^{-5},10^{-3}]$.

The contribution from logical errors of the accumulated depolarizing channel is then
\begin{equation}
    \epsilon_{\mathrm{LER},j} (p)
    =
    \left\|
        \mathcal{D}_1 (q_{\mathrm{acc},j}(p) )
        -
        \mathcal{I}
    \right\|_{\diamond}.
    \label{eq:eps_ler_depol}
\end{equation}
For the depolarizing channel convention in Eq.~\eqref{eq:dep_channel_def}, this evaluates to
$\epsilon_{\mathrm{LER},j}(p) = 2q_{\mathrm{acc},j}(p)$.
Finally, we assign each candidate the total error estimate
$ \epsilon_j (p) = \epsilon_{\mathrm{synth},j} + \epsilon_{\mathrm{LER},j}(p)$.
The query returns the lowest-cost candidate satisfying $\epsilon_j (p) \le \epsilon_{\mathrm{tot}}$.
This additive criterion is conservative: it follows from the triangle inequality for the diamond norm, since the implemented noisy channel differs from the target both because the ideal compiled unitary $\widetilde{U}_j$ only approximates $U$ and because the gates in the sequence accumulate logical noise.

\section{Compilation cost with and without the $\sqT$ gate }\label{section:comparison}

We now use the noisy, cost-aware synthesis framework of Sec.~\ref{section:noisy-database} to compare the two gate sets
\begin{equation}
\gs{1}=\{\mathrm{ H,S,S^\dagger,T,T^\dagger} \}, 
\ 
\gs{2}=\gs{1}\cup\{ \sqT,\sqT^{\dagger}\}.    
\end{equation}
Because each query returns the lowest-cost sequence satisfying the total error budget, the comparison incorporates both synthesis accuracy and the accumulated logical noise of the chosen primitives.

To set the scale of the error budget, we recall that for a $Z$ rotation
\begin{equation}
\left\|\mathcal{R}_{\Z}(\theta)-\mathcal{I}\right\|_{\diamond}
=
2\sin\left(\frac{|\theta|}{2}\right).    
\end{equation}
Thus an error budget $\epsilon$ corresponds to the angular resolution
\begin{equation}
\theta_\epsilon=2\arcsin\left(\frac{\epsilon}{2}\right)\simeq \epsilon    
\end{equation}
for small $\epsilon$. For example, $\epsilon=0.05$ is approximately $\pi/63$, close to the $\pi/64$ reference line used below; similarly, $\pi/128$ and $\pi/256$ correspond to diamond distances of about $2.45\times10^{-2}$ and $1.23\times10^{-2}$.

\subsection{Representative target rotations}
\label{subsection:representative-target-rotations}

We first illustrate the effect of adding $\sqT$ using two representative phase rotations. 
For both examples, we use the same total error budget,
\begin{equation}
    \epsilon_{\mathrm{tot}} = 2\times 10^{-2},
\end{equation}
and evaluate the noisy query at physical error rate $p=10^{-4}$. 
This choice corresponds to approximately degree-scale angular resolution. 

\textit{Example 1:} The largest possible improvement occurs when the target gate is itself $\sqT$, or equivalently $R_{\Z}(\pi/8)$ up to a global phase. 
This gate is native in $\gs{2}$ but must be synthesized over $\gs{1}$. 
At physical error rate $p=10^{-4}$, the best $\gs{1}$ candidate returned by the query has normalized cost $18.40$ and uses $18$ $\T$ injections:
{\small
\begin{align}
\widetilde{U}_{\pi/8}^{(\gs{1})}
&=
\mathrm{H T H T^\dagger H T^\dagger H T^\dagger H T^\dagger
H T^\dagger H T H T H T H}
\notag\\
&\quad
\mathrm{T H T^\dagger H T^\dagger H T^\dagger H T H T^\dagger
H T H T H T H X .}
\label{eq:sqrtT_gs1_decomp}
\end{align}
}
The corresponding error contributions are
\begin{equation}
    \epsilon_{\mathrm{synth}}
    \simeq
    9.9\times 10^{-3},
    \qquad
    \epsilon_{\mathrm{LER}}
    \simeq
    3.5\times 10^{-4},
\end{equation}
so that
\begin{equation}
    \epsilon_{\mathrm{synth}}
    +
    \epsilon_{\mathrm{LER}}
    \simeq
    1.0\times 10^{-2}.
\end{equation}
At this physical error rate, the synthesis error is almost $30$ times
larger than the accumulated logical error and therefore dominates the
total error.
Thus, although the target gate can be approximated to roughly degree-scale accuracy using only Clifford$+ \T$, doing so requires a relatively long sequence of non-Clifford injections.

For comparison, under the same physical error rate and total error budget, $\gs{2}$ implements the target natively using a single $\sqT$ injection. 
The synthesis error is exactly zero, and the only contribution to the total error comes from the noisy logical implementation,
\begin{equation}
    \epsilon_{\mathrm{synth}}=0,
    \qquad
    \epsilon_{\mathrm{LER}}
    \simeq
    3.5\times 10^{-5}.
\end{equation}
The normalized cost is $2.45$. 
Thus, for this target gate, replacing the lengthy Clifford$+ \T$ decomposition of normalized cost $18.40$ by a single $\sqT$ injection of normalized cost $2.45$ reduces the cost by $1-2.45/18.40=86.7\%$.

\textit{Example 2:} As a second example, we consider the smaller phase rotation $R_{\Z}(\pi/16)=\sqrt{\sqT}$ (up to a global phase), which is not natively available in either gate set.
For $\gs{1}$, the query returns a sequence with 17 $\T$ injections,
{\small
\begin{align}
\widetilde{U}_{\pi/16}^{(\gs{1})}
&=
\mathrm{ H T^\dagger H T H T H T^\dagger H T^\dagger H T^\dagger H T H T H}
\notag\\
&\quad
\mathrm{ T H T H T H T^\dagger H T^\dagger H T^\dagger H T H T H T^\dagger H .}
\label{eq:sqrtsqrtT_gs1_decomp}
\end{align}
}
This candidate has normalized cost $17.36$ and
\begin{equation}
    \epsilon_{\mathrm{synth}}
    \simeq
    1.3\times 10^{-2},
    \qquad
    \epsilon_{\mathrm{LER}}
    \simeq
    3.3\times 10^{-4}.
\end{equation}
The total error is therefore approximately $1.33\times 10^{-2}$, which is below the budget.

For the same target  physical error rate and total error budget, the $\gs{2}$ query returns a decomposition using four $\T$ injections and four $\sqT$ injections,
{\small
\begin{equation}
\widetilde{U}_{\pi/16}^{(\gs{2})}
=
\mathrm{ T H \sqrt{T}^{\dagger} S H \sqrt{T}^{\dagger} H T H T H T^\dagger H
\sqrt{T}^{\dagger} H \sqrt{T} S^\dagger H .}
\label{eq:sqrtsqrtT_gs2_decomp}
\end{equation}
}
Its normalized cost is $14.00$, with
\begin{equation}
    \epsilon_{\mathrm{synth}}
    \simeq
    1.6\times 10^{-2},
    \qquad
    \epsilon_{\mathrm{LER}}
    \simeq
    2.2\times 10^{-4}.
\end{equation}
Although this candidate has a slightly larger synthesis error than the $\gs{1}$ sequence, it still satisfies the same total error budget. 
The reduced number of costly non-Clifford injections lowers the normalized cost from $17.36$ to $14.00$, corresponding to a relative
reduction of $1-14.00/17.36\simeq19.4\%$.

These two examples illustrate two complementary mechanisms by which $\sqT$ improves compilation. 
For $R_{\Z}(\pi/8)$, the target is made native, eliminating synthesis error entirely and reducing the cost by almost an order of magnitude. 
For $R_{\Z}(\pi/16)$, the target is not native, but the additional $\sqT$ primitive gives the noisy query more flexibility: it can choose a sequence with slightly larger synthesis error but lower accumulated implementation cost, while still satisfying the same total error budget.

\subsection{Haar-random single-qubit benchmarks}

To test whether the observed savings persist beyond hand-picked examples of phase rotations, we benchmark the two gate sets on $10^4$ Haar-random~\cite{nielsen_chuang} single-qubit unitaries. 
For each target unitary and total error budget $\epsilon_{\mathrm{tot}}$, we query the corresponding database and select the lowest-cost sequence satisfying the total error budget.

The same set of Haar-random targets is used for $\gs{1}$ and $\gs{2}$, so the resulting cost reductions are evaluated as paired comparisons on identical instances. Because the databases are finite, some targets may have no admissible sequence at a given error budget. We therefore report not only the average cost over successful queries, but also the fraction of feasible queries, defined as the fraction of targets for which at least one sequence satisfies the noisy constraint.

\begin{figure}
    \centering
    \includegraphics[width=\columnwidth]{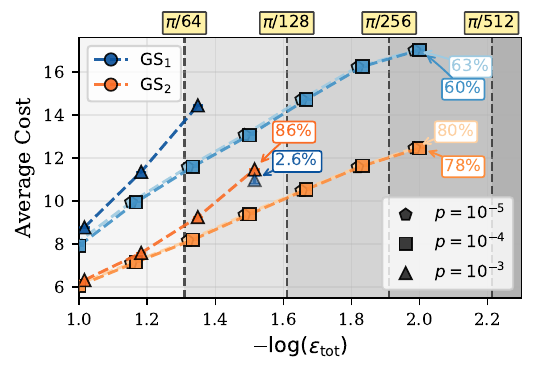}
    \caption{
Average normalized cost of decomposing $10^4$ Haar-random single-qubit unitaries under a total diamond-distance budget $\epsilon_{\mathrm{tot}}$.
Blue data correspond to $\gs{1}=\{ \mathrm{ H,S,S^\dagger,T,T^\dagger} \}$, while orange data correspond to $\gs{2}=\gs{1}\cup\{\sqT,\sqT^{\dagger}\}$.
Different markers denote different physical error rates.
Annotated percentages indicate fraction of feasible queriess below $90\%$; unannotated non-faded points have fraction of feasible queries well above $90\%$.
The faded $\gs{1}$ point at $p=10^{-3}$ is included only as a diagnostic.
Vertical dashed lines indicate the diamond distance between $R_{\Z}(\theta)$ and the identity for $\theta\in\{\pi/64,\pi/128,\pi/256,\pi/512\}$.
Dashed lines connecting data points are simple interpolations included as guide to the eye.
Error bars denote the standard error of the mean over the feasible Haar-random targets.
}
    \label{fig:decomposition_comparison}
\end{figure}

Fig.~\ref{fig:decomposition_comparison} summarizes the Haar-random benchmark. 
It reports the average normalized cost of the lowest-cost feasible decomposition as a function of the total error budget $\epsilon_{\mathrm{tot}}$, with blue points corresponding to $\gs{1}$ and orange points to $\gs{2}$. 
Different marker symbols indicate different physical error rates $p$. 
The vertical dashed lines mark the diamond distance between $R_{\Z}(\theta)$ and the identity for $\theta\in\{\pi/64,\pi/128,\pi/256,\pi/512\}$, providing a reference scale for the angular resolution associated with each error budget. 
All non-faded points shown have fraction of feasible queries at least $60\%$; whenever this fraction is below $90\%$, it is written explicitly next to the point. 
Unannotated points therefore have fraction of feasible queries well above $90\%$. 
The faded $\gs{1}$ point is shown only as a diagnostic and is discussed below.

The figure shows two related advantages of adding $\sqT$ to the native gate set. 
First, $\gs{2}$ consistently lowers the average normalized cost throughout the regime where the query has broad coverage. 
Second, $\gs{2}$ improves coverage at the same cost: the finite $\gs{2}$ database finds admissible decompositions for a larger fraction of Haar-random targets under the same noisy constraint. 
This is important because the plotted averages are conditional on successful queries, so cost reductions are most meaningful when the fraction of feasible queries remains high.

The improvement is already visible in the synthesis-error dominated and weakly noise-aware regimes, especially for $p=10^{-5}$ and $p=10^{-4}$. 
At these physical error rates, the accumulated contribution from logical errors is small compared with the total budget over the displayed range, so the query is dominated primarily by $\epsilon_{\mathrm{synth}}$. 
Consequently, the selected decompositions for $p=10^{-5}$ and $p=10^{-4}$ are nearly identical: the database is effectively choosing the lowest-cost sequence whose noiseless approximation error lies below $\epsilon_{\mathrm{tot}}$.

A representative scale estimate illustrates this point. 
For $\gs{1}$ near $\epsilon_{\mathrm{tot}}=10^{-1.4}\simeq 3.98\times 10^{-2}$, the average normalized cost is approximately $12$, corresponding roughly to eleven $T$ injections and ten Clifford layers. 
Using the depolarizing coefficients in Tab.~\ref{tab:primitive-noise-costs}, the accumulated depolarizing parameter is
\begin{equation}
    q_{\mathrm{acc}}
    =
    \left(10\cdot 21 + 11\cdot 954\right)p^2
    =
    10704p^2 .
\end{equation}
At $p=10^{-4}$ this gives
\begin{equation}
    q_{\mathrm{acc}} \simeq 1.07\times 10^{-4},
    \qquad
    \epsilon_{\mathrm{LER}} = 2q_{\mathrm{acc}}
    \simeq 2.14\times 10^{-4},
\end{equation}
which is only about $0.54\%$ of the total budget. 
Thus, in this regime the error budget is overwhelmingly consumed by synthesis error rather than logical noise.

The $p=10^{-3}$ data show the onset of a different regime. 
For the same representative sequence,
\begin{equation}
    q_{\mathrm{acc}}\simeq 1.07\times 10^{-2},
    \qquad
    \epsilon_{\mathrm{LER}}\simeq 2.14\times 10^{-2},
\end{equation}
so logical noise already consumes about $54\%$ of the budget at $\epsilon_{\mathrm{tot}}=10^{-1.4}$. 
As the target accuracy is tightened, longer decompositions that reduce $\epsilon_{\mathrm{synth}}$ can become infeasible because their accumulated logical error alone uses too much of the allowed error budget. 
The query therefore balances approximation accuracy against the additional logical noise incurred by adding gates.

This feasibility effect is most clearly visible near $\epsilon_{\mathrm{tot}}=10^{-1.5}$ for $p=10^{-3}$. 
With the $\gs{1}$ database restricted to sequences of cost $C\leq 17$, feasible decompositions are found for only $2.6\%$ of the Haar-random targets, as indicated by the faded blue marker. 
This point is not included as part of the main $\gs{1}$ cost trend, because its average is conditioned on too small a subset of targets to be representative. 
By contrast, at the same physical error rate and total error budget, $\gs{2}$ has a fraction of feasible queries of about $86\%$. 
Thus, adding $\sqT$ not only reduces the average cost of successful decompositions, but also substantially increases the coverage of the noisy synthesis procedure within the explored cost regime.

A similar, though less pronounced, coverage advantage appears at lower physical error rates when the target precision is pushed beyond the main plotted range. 
For example, at $p=10^{-4}$ and $\epsilon_{\mathrm{tot}}=10^{-2.2}$, the fraction of feasible queries is $24.6\%$ for $\gs{1}$ and $37.2\%$ for $\gs{2}$. 
At $\epsilon_{\mathrm{tot}}=10^{-2.8}$, these values drop to $0.2\%$ and $0.4\%$, respectively. 
Although both databases lose coverage rapidly in this finite-size regime, $\gs{2}$ maintains a visible coverage advantage.

Finally, the flattening or disappearance of curves at the smallest error budgets should not be interpreted as an asymptotic statement about high-precision synthesis. 
It reflects two effects of the finite resources used in this study: the databases are generated only up to finite maximum cost, and reducing $\epsilon_{\mathrm{synth}}$ by adding more gates also increases $\epsilon_{\mathrm{LER}}$. 
This tradeoff is negligible at $p=10^{-5}$ and remains small at $p=10^{-4}$ over the displayed range, but it becomes dominant at $p=10^{-3}$. 
Thus, the stable comparison between $p=10^{-5}$ and $p=10^{-4}$ captures the robust cost advantage of including $\sqT$, while the higher-$p$ data show the onset of logical-noise-limited synthesis and the associated loss of feasible coverage.

\section{Conclusion}
In this work, we have studied flag-FT preparation of the logical $\ket{\sqT}$ magic state using the tetrahedral color code and its morphed variant. 
The preparation circuits combine a non-FT encoding step, a flagged measurement of the logical $\mathrm{H}_{\sqT}=\T \X$ operator, and a reduced set of stabilizer measurements chosen to detect the remaining dangerous single-fault events. 
Under circuit-level depolarizing noise without idling errors, the logical error rates (LERs) of the accepted state preparation attempts exhibit the expected quadratic suppression in the physical error rate. 
The comparison shows that greater code distance does not necessarily lead to better performance for state preparation: the additional fault locations in the larger tetrahedral-code circuit outweigh its stronger code protection. 
Conversely, the smaller morphed-code protocol achieves a preparation LER close to the Steane code baseline, highlighting the importance of minimizing circuit volume in post-selected FT protocols.

To connect state preparation with compilation, we have characterized the injected logical operations using simulations using six logical input states and quantum process tomography. 
From the reconstructed Pauli-transfer matrices, we have extracted Pauli channel approximations and replaced them by conservative depolarizing upper bounds. 
These logical noise estimates have been incorporated into the synthesis database through a second-order estimate of the accumulated depolarizing error, allowing us to compare Clifford$+ \T$ and Clifford$+ \T + \sqT$ under a common total error budget,
\begin{equation*}
    \epsilon_{\mathrm{synth}}+\epsilon_{\mathrm{LER}}\leq \epsilon_{\mathrm{tot}} .
\end{equation*}
Using this noisy, cost-aware synthesis procedure, our main finding is that adding $\sqrt{T}$ can substantially reduce the expected space-time cost of single-qubit compilation. 
The largest reduction occurs for target unitaries which become native by the additional $\sqT$ primitive, because a direct injection replaces a lengthy Clifford$+ \T$ decomposition.
For non-native target operations, mixed Clifford$+ \T$$+ \sqT$ decompositions can still reduce the cost by requiring fewer costly non-Clifford injections.

For $10^4$ Haar-random single-qubit unitaries, adding $\sqT$ reduces the average chosen cost by approximately $25$--$30\%$ across the error budgets shown in Fig.~\ref{fig:decomposition_comparison}. 
This reduction is stable between the synthesis-dominated regime at $p=10^{-5}$ and the noise-aware regime at $p=10^{-4}$, indicating that the advantage persists after including logical noise in the database query.

Our results open several directions for future work.
First, the flag-based preparation strategy may extend to magic states
for smaller-angle $\Z$ rotations, for example states of the form $R_{\Z}(\pi/2^l)\ket{+}$, motivated by transversal phase gates in Reed--Muller-type constructions~\cite{landahl_2013, duclos_poulin_2015}. 
Second, the preparation and compilation trade-offs studied in this work could be refined further by evaluating them under hardware-specific noise models, including biased noise, leakage, crosstalk, connectivity constraints, and idling errors. 
The moderately small number of physical qubits required in the present circuit constructions also render the protocols suitable for near-term experimental demonstrations in state-of-the-art quantum processors.
Third, the compilation analysis could also be refined further by using the reconstructed anisotropic Pauli channels directly, rather than replacing each primitive by a depolarizing upper bound. 
Finally, it would be interesting to study whether native $\sqT$ resources provide similar savings for multi-qubit compilation, structured algorithmic subroutines, or after further logical-error reduction through cultivation~\cite{gidney_cultivation,amy_2024, Chen2026Jun} or an extension of the protocols developed in this work to larger-distance 3D color codes.

\section*{Software}

We have used Python~\cite{python} for all coding-related tasks,  Numpy~\cite{numpy}, pandas~\cite{pandas}, and Scipy~\cite{scipy} for numerical calculations and Matplotlib~\cite{matplotlib} for visualization.
We have used Cirq~\cite{CIRQ} and Clifft~\cite{clifft} to simulate the quantum circuits under the circuit-level noise model we have considered here.

\section*{Data and code availability}
All the data and the code used to generate the presented results and figures are available in Zenodo~\cite{zenodo_repository}.

\section*{Conflicts of interest}
The authors have no conflicts of interest.

\section*{Author contributions}
B.Y. formulated the theory, conducted the simulations and generated the plots under supervision of M.R.. B.Y. wrote the manuscript with input from M.R. and M.M.. 
M.M. contributed key insights and feedback on both the original motivation and the results.

\section*{Acknowledgements}

We acknowledge support from the Deutsche Forschungsgemeinschaft (DFG, German Research Foundation) under Germany’s Excellence Strategy Cluster of Excellence Matter and Light for Quantum Computing (ML4Q) EXC 2004/1 390534769 and through the DFG Priority Programme SPP
2514. Furthermore, B.Y. and M.M. acknowledge support by the European Union’s Horizon Europe research and innovation programme under Grant Agreement No. 101114305 (“MILLENION-SGA1” EU Project). This research is also part of the Munich Quantum Valley (K-8), which is supported by the Bavarian state government with funds from the Hightech Agenda Bayern Plus. We furthermore acknowledge support by the German Federal Ministry of Research, Technology and Space
(BMFTR) through the project MUNIQC-Atoms (Grant
No. 13N16070) and as part of the Research Program Quantum
Systems, research project 13N17317 (“SQale”). 
The authors acknowledge the use of large language model, in particular ChatGPT 5.6 Sol, to generate the code for the simulations.

\bibliography{references}

\appendix

\section{Additional Pauli Channel Fits}
\label{app:ptm-pauli}

This appendix reports the Pauli channel probabilities extracted for the transversal logical $\Sg$ gate and the logical $\T$-injection circuit using the procedure described in Sec.~\ref{subsection:qpt-ptm}.

\subsection{$S$ Gate}
\label{app:s-ptm-pauli}

Figure~\ref{fig:app-qpt-pauli}(a) shows the Pauli channel probabilities extracted for the transversal logical $S$ gate.
Their quadratic scaling determines the conservative logical-noise coefficient assigned to transversal Clifford operations in Tab.~\ref{tab:primitive-noise-costs}.
The same analysis and observations apply to the other transversal Clifford operations.

\subsection{$T$ Gate}
\label{app:t-ptm-pauli}

\begin{figure}[!b]
\centering
\includegraphics[width=0.95\columnwidth]
{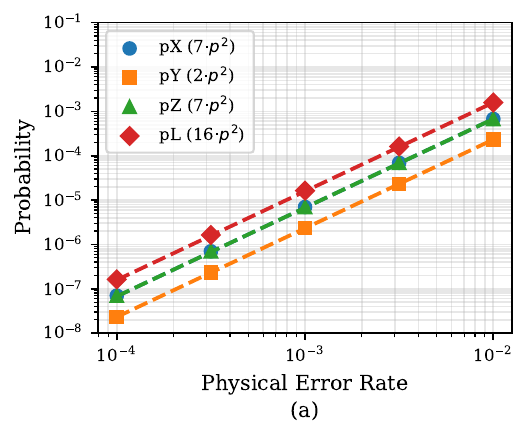}
\par\vspace{0.3em}
\includegraphics[width=0.95\columnwidth]
{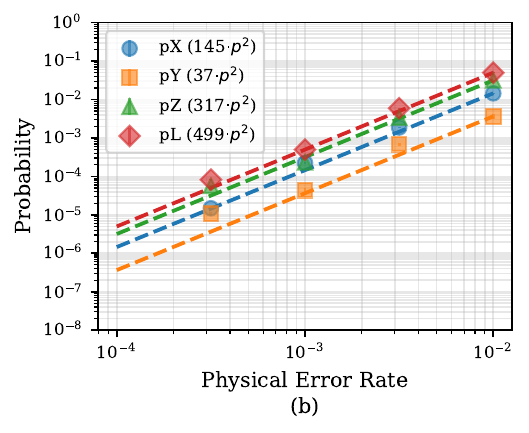}
\caption{
Pauli channel probabilities extracted from the residual PTMs of (a) the transversal logical $\Sg$ gate and (b) the logical $\T$-injection circuit shown in Fig.~\ref{fig:injection}(a).
The points show the probabilities obtained from the simulated data using the diagonal entries of the residual Pauli block according to Eq.~\eqref{eq:pauli_probs_from_ptm}, while the dashed lines show quadratic fits.
For both operations, we obtain the conservative depolarizing error rate from the largest fitted Pauli component according to Eq.~\eqref{eq:depol_upper_bound_rule}.
The resulting coefficients are used for the transversal Clifford gates and the $\T$ injections, respectively, in Tab.~\ref{tab:primitive-noise-costs}.
}
\label{fig:app-qpt-pauli}
\end{figure}

We next apply the same reconstruction procedure to the full logical $T$-injection circuit. 
The extracted Pauli channel probabilities are shown in Fig.~\ref{fig:app-qpt-pauli}(b), and their quadratic fits determine the conservative depolarizing-noise coefficient used for $\T$ injections in Tab.~\ref{tab:primitive-noise-costs}.

\end{document}